\documentstyle[aps,epsf,multicol,amsmath,amsfonts,graphicx,bbm]{revtex}

\begin{document}

\draft

\title{Concepts of a quantum information theory of many letters}
\author{Kim J. Bostr\"om}
\address{
}
\author{}
\address{Institut f\"ur Physik, Universit\"at Potsdam, 14469 Potsdam, Germany}
\date{Split version 2.1 beta / \today}

\maketitle

\begin{abstract}

A theoretical framework is presented allowing the treatment of quantum messages with components of variable length. To this aim a many-letter space, similiar to the Fock space, is constructed, generalizing the standard quantum information theory of block messages of fixed length. In the many-letter space a length operator can be defined measuring the length of a quantum message, whose eigenspaces are the block Hilbert spaces used in the standard theory. 

\end{abstract}

\begin{multicols}{2}
\narrowtext

\section{Introduction}

Information theory is the theory of messages composed from letters. In classical information theory a message is represented by the state of a classical system composed of many subsystems representing the letters of the message. Quantum information theory is much the same, though here the systems are  \emph{quantum}. Since quantum systems obey the laws of quantum mechanics, the situation is radically different from the classical case. However, whereas in classical information theory there is no difficulty in dealing with messages of variable length, quantum information theory, which is usually based on Hilbert spaces of a fixed dimension, does not allow a simple treatment of quantum messages in a superposition of distinct lengths. In this paper a theoretical framework is presented which allows the treatment of such quantum messages in an intuitive way. It is based on the notion of an infinite \emph{direct sum} of Hilbert spaces, which obtains physical meaning if one imagines e.g. a source of photons whose number is a quantum mechanical observable, i.e. the state of the photon ray is generally in a \emph{superposition} of distinct photon number states. The only difference between particles and quantum letters is that the letter systems are \emph{distinguishable}. So the many-letter quantum information theory presented here is just many-particle quantum mechanics with distinguishable particles. Within this framework a close analogy between concepts of classical and quantum information theory can be established, while the standard quantum information theory is fully contained. 

This paper is separated into two parts. The first part reviews roughly some basic concepts of classical information theory in order to motivate the corresponding notions presented in the second part, which is dedicated to quantum information theory. A detailed summary of classical information theory can be found in~\cite{MacKay}, a very recommendable review on quantum information theory is given in~\cite{Preskill}.

\section{Classical messages}

\subsection{General messages and block messages}

The basic object in information theory is a message. A \emph{classical message} is a string $\boldsymbol x$ of letters $x$ taken from an alphabet ${\cal A}$ of size $|{\cal A}|$ and is denoted by
$\boldsymbol x=(x_1\cdots x_n)$.
Let us denote strings of length $n$ explicitely by
\begin{equation}
	x^n:=(x_1\cdots x_n)\quad.
\end{equation}
The set of \emph{block messages} $x^N$ of fixed length $N$ is written as
\begin{equation}
	{\cal A}^N:=\{(x_1\cdots x_N)\mid x_n\in{\cal A}\}\quad.
\end{equation}
Let us also allow for the \emph{empty message} $x^0=(\cdot)$ that forms the set
${\cal A}^0:=\{(\cdot)\}$.
The set of all messages of finite length is defined by
\begin{equation}
	{\cal A}^+:=\bigcup_{n=0}^\infty{\cal A}^n\quad.
\end{equation}
Now Alice wants to communicate general messages to Bob. There are certain messages she wants to send and others (perhaps nonsense or too nasty messages) she does not. So she extracts a \emph{source set} $\Omega\subset{\cal A}^+$ and sends each message $x^n\in\Omega$ with \emph{a priori} probability $p(x^n)>0$. To Bob, who does not know what Alice is doing, the message appears as a random variable $\boldsymbol X$, defined by the source set $\Omega$ and the \emph{a priori} probabilities $p(x^n)$:
\begin{equation}
	\boldsymbol X:=\{[x^n,p(x^n)]\mid x^n\in
	\Omega\}\quad,
\end{equation}
where $p(x^n)>0$ for all $x^n\in\Omega$ and $\sum_{x^n\in\Omega}p(x^n)=1$. The random variable $\boldsymbol X$ is called a \emph{message ensemble}.

If Bob wants to analyze the messages, he performs a measurement on a received message $x^n$ and obtains a real number $A(x^n)$. The ensemble average of his \emph{observable} $A:\Omega\rightarrow\mathbbm R$ is then given by
\begin{equation}
	\overline A(\boldsymbol X)\equiv\,<A(\boldsymbol X)>\,
	:=\sum_{x^n\in\Omega}
	p(x^n)A(x^n)\quad.
\end{equation}
He may, for example, measure the length of a message, given by the \emph{length function} $L:{\cal A}^+\rightarrow{\mathbbm N}$ with $L(x^n)=n$ and the length of the empty message being set to $L(\cdot):=0$. The expected length of a message from Alice is then
\begin{equation}
	\overline L(\boldsymbol X)=\sum_{x^n\in\Omega}p(x^n)\,n\quad.
\end{equation}
If $\boldsymbol X$ is a block message, its length is fixed to some $N$ that is known to both Alice and Bob.

\subsection{Canonical messages}

There is a type of message that is of fundamental importance to information theory and that is why it is named here the \emph{canonical message}. It is a block message of fixed length $N$ formed by independent identically distributed letters. Alice takes the \emph{letter ensemble}
\begin{equation}
	X:=\{[x,p(x)]\mid x\in{\cal A}\}
\end{equation}
and composes messages by just putting $N$ letters in a row, i.e.
$\boldsymbol X=X^N:=(X_1\cdots X_N),X_n=X$, 
resulting in a \emph{canonical message ensemble}
\begin{equation}
	X^N=\{[x^N,p(x^N)]\mid x^N\in{\cal A}^N\}
\end{equation}
with $p(x^N)=p(x_1)\cdots p(x_N)$.

\section{Quantum messages}

Quantum information theory can be obtained straightforwardly by mapping classical objects to quantum objects. To put it simple: \emph{Classical information} is carried by \emph{classical states} of a medium and \emph{quantum information} is carried by \emph{quantum states} of a medium. Imagine Alice writing her messages not on a sheet of paper or imprinting it onto the surface  of a magnetic tape or a hard disk, but instead modifying single atoms, molecules, electrons, photons or any other microscopic systems that can only be described by the laws of quantum mechanics. The mathematical framework of classical information theory then has to be translated into the language of quantum mechanics. The result of this quantization procedure is \emph{quantum information theory}.

Alice prepares the medium to be in a quantum state $|\varphi\rangle$, performs some operations on it and sends it to Bob. The message has been successfully transmitted, if Bob, after performing some operations on the received state, ends up with the same state $|\varphi\rangle$ that Alice originally prepared. Note that it is not necessary for Bob to perform any \emph{measurement} on the state. Bob does not need to \emph{know} which state Alice has originally prepared. This would transform the quantum information contained in that state to classical information. It is a major difference between classical and quantum information that \emph{knowledge}, i.e. the state of someone's brain, is always classical, whereas the state of an unknown quantum state is \emph{intrinsically unknowable}, since there is no operation in the world allowing to guess an unknown state with perfect fidelity.

\subsection{Quantum alphabet}

\subsubsection{A priori alphabet}

A classical letter is represented by the state of a classical system. If the system is quantum instead, the letter corresponds to a quantum state. Thus a classical letter $x$ can be transformed into a quantum letter by mapping it to a normalized Hilbert vector $|x\rangle\in{\cal H}$. In such a way, the classical alphabet ${\cal A}$ is mapped to a \emph{quantum alphabet}
\begin{equation}
	{\cal Q}:=\{|x\rangle\in{\cal H}\mid x\in{\cal A}\}\quad.
\end{equation}
The Hilbert space spanned by the letters of the quantum alphabet is the \emph{letter space}
\begin{equation}
	{\cal H}_{\cal Q}:=\text{Span}({\cal Q})\quad,
\end{equation}
where its dimension given by
$K_{\cal Q}:=\dim{\cal H}_{\cal Q}\leq|{\cal Q}|=|{\cal A}|$, 
with equality if the letter states are linearly independent.

The quantum letters in ${\cal Q}$ are not required to be mutual orthogonal, yet not even linearly independent. So it is in general not possible for Bob to perfectly \emph{distinguish} the letters that Alice choses from her alphabet ${\cal Q}$, which is thus called an \emph{a priori alphabet}. In sad words: Bob will generally not be able to \emph{read} a message from Alice. Instead he probably recognizes \emph{a posteriori} letters, that also lie in the letter space spanned by the \emph{a priori} alphabet but which are different from the letters that Alice originally had sent. This is a typically quantum phenomenon with no classical correspondance.

\subsubsection{Basis alphabet}

A set ${\cal B}_{\cal Q}=\{|a\rangle\}_a$ of mutually orthogonal normalized basis vectors of the letter space ${\cal H}_{\cal Q}$ is called a \emph{basis alphabet} corresponding to ${\cal Q}$, so
\begin{equation}
	\sum_{a\in{\cal B}_{\cal Q}}|a\rangle\langle a|
	={\mathbbm1}_{{\cal H}_{\cal Q}}\quad
	\langle a|a'\rangle=\delta_{aa'}\quad.
\end{equation}	
Since all \emph{basis letters} $|a\rangle\in{\cal B}_{\cal Q}$ are perfectly distinguishable, they can be viewed as almost classical. The basis alphabet is a very important concept in quantum information theory. Any single-letter message from Alice can be expressed as a superposition of basis letters:
\begin{equation}
	|x\rangle=\sum_{a\in{\cal B}_{\cal Q}}\langle a|x\rangle |a\rangle\quad.
\end{equation} 
If Bob happens to measure along the basis letter subspaces, the \emph{a priori} message from Alice will decohere into its basis letter components. Up to the measurement, though, they are all simultaneously engaged.
The number of basis letters equals the dimension of the letter space, so there are probably less basis letters than \emph{a priori} letters, i.e.
$|{\cal B}_{\cal Q}|=\dim{\cal H}_{\cal Q}\leq|{\cal Q}|$.

\subsection{Block messages}

\subsubsection{General block messages}

A classical string $x^n\in{\cal A}^n$ of length $n$, given by  $x^n=(x_1\cdots x_n)$, is mapped to a Hilbert vector $|x^n\rangle\in{\cal H}^n$ normalized to unity and formed by the tensor product of the letter states corresponding to the letters contained in $x^n$. It is called a \emph{product message} or \emph{quantum string} and is denoted by
\begin{equation}
	|x^n\rangle:=|x_1\cdots x_n\rangle
	\equiv|x_1\rangle\otimes\cdots\otimes|x_n\rangle\quad.
\end{equation}
The set ${\cal A}^N$ of classical strings of fixed length $N$ is mapped to the set of \emph{quantum block strings}
\begin{equation}
	{\cal Q}^N:=\{|x^N\rangle\in{\cal H}^N\mid x^N\in{\cal A}^N\}\quad.
\end{equation}
Let us allow also for the empty quantum message $|x^0\rangle=|\cdot\rangle$ that forms the set
${\cal Q}^0:=\{|\cdot\rangle\}$.
The Hilbert space spanned by the elements of ${\cal Q}^N$ is the $N$-fold tensor product of the letter space and is called the \emph{block message space}:
\begin{equation}
	{\cal H}_{\cal Q}^N:=\text{Span}({\cal Q}^N)
	={\cal H}_{\cal Q}\otimes\cdots
	\otimes{\cal H}_{\cal Q}\quad,
\end{equation}
where its dimension is given by
$\dim{\cal H}_{\cal Q}^N=\left(\dim{\cal H}_{\cal Q}\right)^N\leq |{\cal Q}|^N$.
The one-dimensional empty message space is defined by
${\cal H}_{\cal Q}^0:=\text{Span}({\cal Q}^0)$.

What messages can Alice compose now? She can prepare the quantum string $|x^N\rangle$ of length $N$ by manipulating each of the $N$ letter systems separately. But quantum mechanics allows her also to perform unitary operations on the entire message state $|x^N\rangle\in{\cal Q}^N$ before sending it to Bob. So she can construct any normalized vector $|\varphi(x^N)\rangle\in{\cal H}_{\cal Q}^N$ by performing $|\varphi(x^N)\rangle=U(x^N)|x^N\rangle$,
where $U(x^N)$ is a unitary operator on ${\cal H}_{\cal Q}^N$.
Though $|x^N\rangle$ is a \emph{product message}, $|\varphi(x^N)\rangle$ generally is not. In that case it is an \emph{entangled message}. While quantum strings are always product messages, general block messages can be arbitrary superpositions of strings of the same length. There is no classical correspondance to such objects. In order to make it explicetely we may denote a block message $|\varphi\rangle$ of length $N$ by a small index $N$ like in $|\varphi\rangle_N$. A general form of block messages is then given by  
\begin{equation}
	|\varphi\rangle_N=\sum_{a^N}\varphi(a^N)|a^N\rangle\quad,
\end{equation}
where $\varphi(a^N)=\langle a^N|\varphi\rangle$ and ${\cal B}_{\cal Q}=\{a\}_a$ being a set of mutually orthogonal basis letters of the letter space ${\cal H}_{\cal Q}$. The sum is performed over all strings $|a^N\rangle$ over the basis alphabet. By applying her unitary operations to the strings $|x^N\rangle$ of ${\cal Q}^N$, Alice prepares a set of general \emph{block messages} of fixed length $N$
\begin{equation}
	 \Gamma:=\{|\varphi\rangle_N\in{\cal H}_{\cal Q}^N
	 \mid p(\varphi)>0\}\quad.
\end{equation}
Alice choses the message $|\varphi\rangle_N\in\Gamma$ with \emph{a priori} probability $p(\varphi)$, i.e. she draws each message from the \emph{message ensemble}
\begin{equation}
	|\Phi\rangle_N:=\{[|\varphi\rangle_N,p(\varphi)]
	\mid |\varphi\rangle_N\in\Gamma\}\quad.
\end{equation}
If Bob receives the message $|\varphi\rangle_N$ and tries to get some classical information out of it, he performs a measurement of an observable $A$, represented by a self-adjoint operator on ${\cal H}_{\cal Q}^N$. Each time he does, he gets a random result whose quantum mechanical expectation value is given by
\begin{equation}
	A(\varphi)\equiv\,<A>_{\varphi}:=\langle\varphi|A|\varphi\rangle_N\quad.
\end{equation}
Since Alice draws her messages from the ensemble $|\Phi\rangle_N$, the \emph{ensemble average} of $A$ is ruled by
\begin{equation}
	A(\Phi):=\,<\langle\Phi|A|\Phi\rangle_N>\,
	=\sum_{\varphi\in\Gamma}p(\varphi)\,
	\langle\varphi|A|\varphi\rangle_N\quad.
\end{equation}
Equivalently, Bob can calculate the ensemble average using the \emph{message matrix}
\begin{equation}
	\sigma:=\sum_{\varphi\in\Gamma}p(\varphi)
	|\varphi\rangle\langle\varphi|_N
\end{equation}
and find the ensemble average being governed by
\begin{equation}
	A(\sigma)\equiv\,<A>_\sigma\,:=\text{Tr}_N\{\sigma A\}\quad,
\end{equation}
where $\text{Tr}_N$ denotes the trace over the space ${\cal H}_{\cal Q}^N$.
It is a profound peculiarity of quantum mechanics that Bob would end up with the same statistical average, if Alice had taken some other message ensemble yielding the same message matrix $\sigma$. Consequently, there is more information in knowing the \emph{ensemble} $|\Phi\rangle$ (like Alice does) than just knowing the \emph{matrix} $\sigma$. This additional information is in no way available by performing measurements on the message states. Nevertheless, these two distinct notions are both used within quantum information theory.

To Bob there would be no difference if Alice had taken the message ensemble
\begin{equation}
	|E\rangle_N:=\{\big[|e_k\rangle_N,q_k\big]\mid k=1\ldots K^N\}\quad,
\end{equation}
with the $|e_k\rangle_N$'s being the eigenstates of $\sigma$,
\begin{equation}
	\sigma=\sum_{k=1}^{K^N}q_k |e_k\rangle\langle e_k|_N\quad,
\end{equation}
where
\begin{equation}
	\langle e_k|e_l\rangle_N=\delta_{kl},\quad
	\sum_{k=1}^{K^N}  |e_k\rangle\langle e_k|_N={\mathbbm1}_N
\end{equation}
and $K^N:=\dim{\cal H}_{\cal Q}^N=(\dim{\cal H}_{\cal Q})^N$. 
This special ensemble is a very interesting one, since here the single messages $|e_k\rangle_N$ can be \emph{distinguished} from another by a suitible measurement. So it is the most classical equivalent ensemble corresponding to what Alice is doing.

\subsubsection{Product block messages}

Alice prepares her message letter by letter and obtains a product state $|x^N\rangle=|x_1\cdots x_N\rangle\in{\cal H}_{\cal Q}^N$. She prepares the state $|x^N\rangle$ with \emph{a priori} probability $p(x^N)$, i.e. she draws her messages from the \emph{product message ensemble}
\begin{equation}
	|X^N\rangle:=\{[|x^N\rangle,p(x^N)]
	\mid x^N\in\Omega\subset{\cal A}^N\}\quad.
\end{equation}
Now the corresponding message matrix,
\begin{eqnarray}
	\sigma&=&\sum_{x_1\cdots x_N}p(x_1\cdots x_N)\Big[
	|x_1\rangle\langle x_1|\otimes\cdots\otimes|x_N\rangle\langle x_N|\Big]\\
	&=&\left[\sum_{x_1}p_1(x_1)|x_1\rangle\langle x_1|\right]
	\otimes\cdots\nonumber\\
	&&\cdots\otimes
	\left[\sum_{x_N}p_N(x_N)|x_N\rangle\langle x_N|\right]\quad,
\end{eqnarray}
falls apart into a product $\sigma=\rho_1\otimes\cdots\otimes\rho_N$ of \emph{single-letter matrices} $\rho_n$, given by
\begin{equation}
	\rho_n:=\sum_{x_n\in{\cal A}}p_n(x_n)|x_n\rangle\langle x_n|\quad,
\end{equation}
with the \emph{marginal probabilities}
\begin{equation}
	p_n(x_n):=\sum_{x_i:\, i\neq n}p(x_1\cdots x_N)\quad.	
\end{equation}
Again it is interesting to regard the spectral decomposition of each single-letter matrix,
\begin{equation}
	\rho_n=\sum_{k=1}^K q_{nk}|e_{nk}\rangle
	\langle e_{nk}|\quad,
\end{equation}
with\begin{equation}
	\langle e_{nk}|e_{nl}\rangle=\delta_{kl},\quad
	\sum_{k=1}^{K}  |e_{nk}\rangle\langle e_{nk}|={\mathbbm1}_{{\cal H}_{\cal Q}}
\end{equation}
and $K:=\dim{\cal H}_{\cal Q}$. 
To Bob it appears as if Alice had prepared messages over orthogonal basis alphabets
${\cal B}_n:=\{[|e_{nk}\rangle,q_{nk}]\mid k=1\ldots K\}$, 
that vary from letter to letter.

\subsubsection{Canonical messages}\label{canonical}

Canonical messages are product block messages $|x^N\rangle\in{\cal H}_{\cal Q}^N$ over an \emph{a priori} alphabet ${\cal Q}=\{|x\rangle\}_x$, chosen with factorizing \emph{a priori} probabilities $p(x^N)=p(x_1)\cdots p(x_N)$.
The message matrix,
\begin{equation}
	\sigma=\rho^{\otimes N}=\rho\otimes\cdots\otimes\rho\quad,
\end{equation}
is the $N$-fold tensor product of the \emph{letter matrix},
\begin{equation}
	\rho=\sum_{x}p(x)\,|x\rangle\langle x|\quad.
\end{equation}
Alice uses an \emph{a priori} letter ensemble
\begin{equation}
	|X\rangle=\{[|x\rangle,p(x)]\mid |x\rangle\in{\cal Q}\}\quad,
\end{equation}
which is just put in a row $N$ times to form the \emph{canonical ensemble} $|X^N\rangle$. To Bob there is no difference if Alice instead uses the \emph{basis letter} ensemble $|A\rangle$ consisting of the $\rho$ eigenstates $|a\rangle$, i.e.
\begin{equation}
	\rho=\sum_{a} q(a)|a\rangle\langle a|\quad,
\end{equation}
and forms the message ensemble $|A^N\rangle$.

\section{Many-letter messages}

\subsection{General many-letter messages}

Standard quantum information theory describes only block messages. We like to go further now and allow quantum messages of arbitrary length. To this aim we seek a quantum analog of the set ${\cal A}^+$ of classical messages of arbitrary length. It is easily found by mapping each classical message $x^n\in{\cal A}^+$ to a product Hilbert vector $|x^n\rangle\in{\cal H}^n$. Regard the set of product block messages of length $n$,
\begin{equation}
	{\cal Q}^n:=\{|x^n\rangle\in{\cal H}^n\mid x^n\in{\cal A}^n\}\quad,
\end{equation}
with ${\cal Q}^0:=\{|\cdot\rangle\}$ being defined as the set formed by the empty message $|\cdot\rangle$.
The Hilbert space spanned by the members of ${\cal Q}^n$ is given by
\begin{equation}
	{\cal H}_{\cal Q}^n:=\text{Span}({\cal Q}^n)\quad.
\end{equation}
Now construct the infinite set
\begin{equation}
	{\cal Q}_+:=\bigcup_{n=0}^\infty{\cal Q}^n\quad.
\end{equation}
The space spanned by the elements of ${\cal Q}_+$ (regarding that messages of distinct length are always orthogonal) is the \emph{many-letter space}
\begin{eqnarray}
	{\cal M}_{\cal Q}&:=&\bigoplus_{n=0}^\infty{\cal H}_{\cal Q}^n\quad.
\end{eqnarray}
The \emph{direct sum} of two Hilbert spaces ${\cal H}_1,{\cal H}_2$ is defined as the orthogonal sum of their elements, i.e.
\begin{equation}
	\begin{split}
		{\cal H}_1&\oplus{\cal H}_2\\
		&:=\{|\psi_1\rangle_1+|\psi_2\rangle_2
		\mid |\psi_1\rangle_1\in{\cal H}_1,|\psi_2\rangle_2\in{\cal H}_2\},
	\end{split}
\end{equation}
and is a Hilbert space with the scalar product
\begin{equation}
	\begin{split}
	\Big({}_1\langle\psi_1|+{}_2\langle\psi_2|\Big)&
	\Big(|\varphi_1\rangle_1+|\varphi_2\rangle_2\Big)\\
	&:=\langle\psi_1|\varphi_1\rangle_1+\langle\psi_2|\varphi_2\rangle_2\quad,
	\end{split}
\end{equation}
and the dimension $\dim({\cal H}_1\oplus{\cal H}_2)=\dim{\cal H}_1+\dim{\cal H}_2$. Both spaces ${\cal H}_1$ and ${\cal H}_2$ are orthogonal subspaces of ${\cal H}_1\oplus{\cal H}_2$, i.e. ${\cal H}_1,{\cal H}_2
\subset({\cal H}_1\oplus{\cal H}_2)$ and
\begin{equation}
	{}_1\langle\psi_1|\psi_2\rangle_2=0
	\quad\forall |\psi_1\rangle_1\in{\cal H}_1,|\psi_2\rangle_2\in{\cal H}_2
	\quad.
\end{equation}
In order to simplify the notation the small indices indicating the Hilbert space a particular component belongs to, are left out.

Maybe the notion of a \emph{direct sum} of Hilbert spaces appears rather unphysical to the reader, since everything in quantum mechanics is usually described in terms of direct \emph{products}. But without explicitly using it, the direct sum is always present. For example, the fundamental space of quantum information theory, the space ${\mathbbm C}^2$ of a single qbit is in fact the direct sum of two ${\mathbbm C}$'s. The reason is that the \emph{cartesian product} ${\mathbbm C}^2={\mathbbm C}\times{\mathbbm C}$ can be embedded into the direct sum ${\mathbbm C}\oplus{\mathbbm C}$ by preserving the Hilbert space structure, since both components of ${\mathbbm C}^2$ are mutually othogonal. Hence the cartesian product and the direct sum are just different representations of the same principle: adding separate \emph{levels} of an observable, i.e. combining properties of a system by a quantum mechanical \emph{OR}. In many-letters theory the distinct degenerate levels of the length operator (see section~\ref{length_op}) are added, that is all.

The space ${\cal M}_{\cal Q}$ contains just \emph{any} quantum message that can be composed from the quantum alphabet ${\cal Q}$ by preparing each single letter state separately and then performing a unitary operation in the many-letter space on the entire message.
Every Hilbert space ${\cal H}_{\cal Q}^n$ of block messages is a \emph{subspace} of ${\cal M}_{\cal Q}$, 
\begin{equation}
	\forall n\in{\mathbbm N}:\quad
	{\cal H}_{\cal Q}^n\subset{\cal M}_{\cal Q}\quad,
\end{equation}
such that every Hilbert vector $|\psi\rangle\in{\cal H}_{\cal Q}^n$ is also an element of ${\cal M}_{\cal Q}$. That way, quantum information theory based on many-letter spaces \emph{contains} quantum information theory based on block spaces, it can be viewed as a straight generalization of the latter. 
The many-letter space ${\cal M}_{\cal Q}$ is similiar to the Fock space used in quantum optics or quantum statistics except that the states contained in ${\cal M}_{\cal Q}$ are neither symmetrized nor antisymmetrized. The \emph{ordering} of the subspaces still matters (imagine a book written without ordering of the letters!). Therefore, the elements of ${\cal M}_{\cal Q}$ are neither Fermions nor Bosons, they are simply \emph{quantum letters} and have to be realized by \emph{distinguishable} quantum systems being separated in space or time or differing in some other observable property, such that their mutual overlap is neglectable.

Of course, the many-letter space can be restricted to a maximum number of letters,
\begin{equation}
	{\cal M}_{\cal Q}^N:=\bigoplus_{n=0}^N {\cal H}_{\cal Q}^n\quad,
\end{equation}
due to a finite reservoir of available qbits. It is a subspace of the total many-letter space, ${\cal M}_{\cal Q}^N\subset{\cal M}_{\cal Q}$, and each many-letter message can be truncated to this subspace by the appropriate projector.

Quantum mechanics allows Alice to compose any superposition of block messages into a \emph{general many-letter message}. Thus she uses the general message ensemble
\begin{equation}
	|\Phi\rangle=\{[|\varphi\rangle,p(\varphi)]
	\mid |\varphi\rangle\in\Gamma\}\quad,
\end{equation}
with the \emph{source set} $\Gamma$ of quantum messages being chosen with nonzero \emph{a priori} probability $p(\varphi)$,
\begin{equation}
	\Gamma=\{|\varphi\rangle\in{\cal M}_{\cal Q}
	\mid p(\varphi)>0\}\quad.
\end{equation}
Note that $\Gamma$ may be an infinite set. The subspace spanned by the elements of $\Gamma$ is the \emph{source space} ${\cal M}_\Gamma\subset{\cal M}_{\cal Q}$,
\begin{equation}
	{\cal M}_\Gamma:=\text{Span}(\Gamma)\quad,
\end{equation}
whose dimension $G:=\dim{\cal M}_\Gamma$ may also be infinite.
Equivalently, the message ensemble may be represented by a corresponding \emph{message matrix} $\sigma\in{\cal S}({\cal M}_{\cal Q})$, given by
\begin{equation}
	\sigma=\sum_{\varphi\in\Gamma}p(\varphi)|\varphi\rangle\langle\varphi|\quad.
\end{equation}
It is often more convenient to use the spectral decomposition of $\sigma$ instead, given by
\begin{equation}
	\sigma=\sum_{i=1}^G q_i|e_i\rangle\langle e_i|\quad,
\end{equation}
where the $|e_i\rangle$'s form an orthonormal basis ${\cal B}_\Gamma$ of the source space ${\cal M}_\Gamma$.

\subsection{Length operator}\label{length_op}

To any classical letter $\boldsymbol x\in{\cal A}^+$ there is a length function $L:{\cal A}^+\rightarrow{\mathbbm N}$ mapping each letter $\boldsymbol x$ to its length $L(\boldsymbol x)$. Since the length of a quantum message is also an observable property (Bob has to measure the number of letter systems being engaged), there is a self-adjoint \emph{length operator} $\widehat L$ acting on the many-letter space ${\cal M}_{\cal Q}$ with a spectral decomposition of mutually orthogonal projectors $\Pi_n$ on ${\cal M}_{\cal Q}$, such that
\begin{equation}
	\widehat L=\sum_{n=0}^\infty n\, \Pi_n \quad,
\end{equation}
with
\begin{equation}
	\Pi_n\,\Pi_m=\delta_{nm}\Pi_n,\quad
	\sum_{n=1}^\infty  \Pi_n ={\mathbbm1}\quad.
\end{equation}
The eigenspaces of the length operator are the block message spaces ${\cal H}_{\cal Q}^n$, which are subspaces of the many-letter space ${\cal M}_{\cal Q}$. Hence the eigenvalues of $\widehat L$ are degenerate by
$K^n:=\dim{\cal H}_{\cal Q}^n=(\dim{\cal H}_{\cal Q})^n$.
The projector $\Pi_n$ onto the subspace ${\cal H}_{\cal Q}^n$ can be decomposed into mutually orthogonal product messages $|a^n\rangle$ of length $n$  composed from a basis alphabet ${\cal B}_{\cal Q}=\{|a\rangle\}_a$, where $|{\cal B}_{\cal Q}^n|=|{\cal B}_{\cal Q}|^n=K^n$ and 
$|a^0\rangle:=|\cdot\rangle\in{\cal H}_{\cal Q}^0$.
The set of product messages of length $n$  composed from the basis alphabet ${\cal B}_{\cal Q}$ is denoted by ${\cal B}_{\cal Q}^n:=\{|a^n\rangle\}_{a^n}$, the basis for the one-dimensional empty message space by ${\cal B}_{\cal Q}^0=\{|\cdot\rangle\}$.
So the projector $\Pi_n$ may be decomposed as
\begin{equation}
	 \Pi_n =\sum_{a^n}
	 |a^n\rangle\langle a^n|\quad,
\end{equation}
where we understand the sum as being performed over all quantum strings $|a^n\rangle\in{\cal B}_{\cal Q}^n$ here and in the following.
Using the basis
\begin{equation}
	{\cal B}_{\cal Q}^+:=\bigcup_{n=0}^\infty {\cal B}_{\cal Q}^n\quad,
\end{equation}
one arrives at the unity decomposition
\begin{equation}
	\sum_{n=0}^\infty\sum_{a^n}
	 |a^n\rangle\langle a^n|={\mathbbm1}\quad,
\end{equation}
where the length operator becomes diagonal.
Now Alice choses a \emph{general many-letter message} $|\varphi\rangle\in\Gamma\subset{\cal M}_{\cal Q}$, whose decomposition in the basis ${\cal B}_{\cal Q}^+$ thus reads
\begin{equation}
	|\varphi\rangle=\sum_{n=0}^\infty\sum_{a^n}
	\varphi(a^n)|a^n\rangle\quad,
\end{equation}
with its wave components given by
\begin{equation}
	\varphi(a^n):=\langle a^n|\varphi\rangle\quad.
\end{equation}
She sends her message to Bob using a quantum channel that is protected against decoherence of the basis vectors $|a^n\rangle$. That way, superpositions of these vectors are preserved and Bob receives the same state that Alice prepared. Now he measures the length of the message, obtaining random results with the \emph{expected length} given by
\begin{equation}
	L(\varphi)=\langle\varphi|\widehat L|\varphi\rangle
	=\sum_{n=0}^\infty\sum_{a^n}
	|\varphi(a^n)|^2\,n\quad,
\end{equation} 
whereas the \emph{ensemble length} of the message $\sigma$ is given by
\begin{eqnarray}
	L(\sigma)&=&\,<\langle\Phi|\widehat L|\Phi\rangle>\,
	=\sum_{\varphi\in\Gamma}p(\varphi)\, L(\varphi)\\
	&=&\sum_{\varphi\in\Gamma}
	\sum_{n=0}^\infty\sum_{a^n}
	p(\varphi)\,|\varphi(a^n)|^2\,n\\
	&=&\text{Tr}\{\sigma \widehat L\}\quad.
\end{eqnarray}
As a generalization, we can define the \emph{expected length} of any (pure or mixed) message, represented by a density matrix $\rho\in{\cal S}({\cal M}_{\cal Q})$, by
\begin{equation}
	 L(\rho):=\text{Tr}\{\rho\widehat L\}\quad.
\end{equation}
Needless to say, the measurement of the length of a message will result in losing all quantum correlations between wave components of distinct length.

\subsection{Random block messages}

Alice now choses block messages $|\varphi\rangle$ from any one of the subspaces ${\cal H}_{\cal Q}^n\subset{\cal M}_{\cal Q}$ with \emph{a priori} probabilities $p(\varphi)$, i.e. she draws her messages from the ensemble
\begin{equation}
	|\Phi\rangle=\{[|\varphi\rangle,p(\varphi)]
	\mid|\varphi\rangle\in\Gamma\}\quad,
\end{equation}
where $\Gamma$ is the set of block messages chosen with nonzero probability:
\begin{equation}
	\Gamma:=\{|\varphi\rangle\in{\cal H}_{\cal Q}^n
	\mid p(\varphi)>0,\,n=0,1,2,\ldots\}\quad,
\end{equation}
The correponding message matrix reads
\begin{equation}
	\sigma=\sum_{\varphi\in\Gamma}p(\varphi)
	|\varphi\rangle\langle\varphi|\quad.
\end{equation} 
Every message $|\varphi\rangle$ drawn from the ensemble has a well-defined length $L(\varphi)$ because it is in one of the eigenspaces of the length operator, i.e. $\widehat L|\varphi\rangle=L(\varphi)|\varphi\rangle$.
Thus the message matrix of random block messages can be \emph{block-diagonalized} into the convex combination of \emph{block matrices} $\sigma_n$,
\begin{equation}
	\sigma=\sum_{n=0}^\infty \lambda_n\,\sigma_n\quad,
\end{equation}
with the \emph{length probabilities} $\lambda_n$, given by
\begin{equation}
	\lambda_n:=\sum_{L(\varphi)=n}p(\varphi)\quad,
\end{equation}
such that 
\begin{equation}
	\sum_{n=0}^\infty \lambda_n=\sum_{n=0}^\infty\sum_{L(\varphi)=n}p(\varphi)
	=\sum_{\varphi\in\Gamma}p(\varphi)=1.
\end{equation}
Every block matrix has a definite length $\widehat L\,\sigma_n=n\,\sigma_n$, hence it commutes with the length operator.
So the average length of the ensemble reads
\begin{equation}
	L(\sigma)=\sum_{n=0}^\infty\lambda_n\, n\quad.
\end{equation}
We chose basis sets $B_n$ of mutually orthogonal block messages $|e_{i_n}^n\rangle$ of length $n$, so that the block matrices become diagonal:
\begin{equation}
	\sigma_n=\sum_{L(\varphi)=n}
	\sum_{i_n=1}^{K_n}|\varphi_{i_n}^n|^2|e_{i_n}^n\rangle
	\langle e_{i_n}^n|\quad,
\end{equation}
with the wave components $\varphi_{i_n}^n:=\langle e_{i_n}^n|\varphi\rangle$.
Note that the block messages $|e_{i_n}^n\rangle$ are generally no product messages.

To Bob it appears as if Alice would send him states $|e_{i_n}^n\rangle$ of well-defined length $n$ with the probability
\begin{equation}
	P(e_{i_n}^n)=\lambda_n\sum_{L(\varphi)=n}|\varphi_{i_n}^n|^2\quad.
\end{equation}
A major advantage of using random block messages is that the length may be measured without disturbing the message. 

\subsection{Grand canonical messages}\label{grand_canonical}

\emph{Grand canonical messages} (or \emph{random canonical messages}) are canonical messages of variable length (just like in thermodynamics, where grand canonical ensembles are canonical ensembles with variable particle number). 
Each classical letter $x^n\in{\cal A}^+$ of variable length $n$, composed from a classical alphabet ${\cal A}$ is mapped to a product vector $|x^n\rangle\in{\cal H}_{\cal Q}^n$ and chosen by Alice with \emph{a priori} probability $p(x^n)$. Alice thus draws her quantum messages from the ensemble
\begin{equation}
	|\boldsymbol X\rangle=\{[|x^n\rangle,p(x^n)]
	\mid |x^n\rangle\in\Gamma,\,n=0,1,2,\ldots\},
\end{equation}
where the source set $\Gamma=\{|x^n\rangle\in{\cal H}_{\cal Q}^n
	\mid p(x^n)>0,\,n=0,1,2\ldots\}$, consists of canonical messages $|x^n\rangle$ of variable length $\widehat L|x^n\rangle=n\,|x^n\rangle$, distributed by
\begin{equation}
	p(x^n):=\lambda_n\,p(x_1)\cdots p(x_n)\quad,
\end{equation}
where
\begin{equation}
	\sum_{x}p(x)=1,
	\quad\sum_{n=0}^\infty\lambda_n=1\quad.
\end{equation}
The grand canonical message matrix has the form
\begin{equation}
	\sigma=\sum_{n=0}^\infty\lambda_n\,\rho^{\otimes n}\quad,
\end{equation}
with the \emph{block matrices}
\begin{equation}
	\rho^{\otimes n}=\rho\otimes\cdots\otimes\rho\quad,
\end{equation}
and the \emph{letter matrices}
\begin{equation}
	\rho=\sum_{x}p(x)|x\rangle\langle x|\quad.
\end{equation}
Each block matrix has a definite length $\widehat L\,
\rho^{\otimes n}=n\,\rho^{\otimes n}$, so
the \emph{average length} of a grand canonical message ensemble is given by
\begin{equation}
	L(\sigma)\,=\sum_{n=0}^\infty\lambda_n\, n\quad.
\end{equation}
Grand canonical messages can be viewed as a gerneralization of canonical messages, in that the length of a message is allowed to vary. 
Just as for every random block message, grand canonical messages are not disturbed by measuring the length operator.

We chose the basis sets ${\cal B}_{\cal Q}=\{|a\rangle\}$ so that the letter matrices become diagonal. The message matrix now reads
\begin{equation}
	\sigma=\sum_{n=0}^\infty\sum_{x^n,a^n}
	\lambda_n\,p(x_1)\cdots p(x_n)\,
	|x^n(a^n)|^2\,|a^n\rangle\langle a^n|,
\end{equation}
with the wave components 
$x^n(a^n):=\langle a^n|x^n\rangle
=\langle a_1|x_1\rangle\cdots\langle a_n|x_n\rangle$.
To Bob it appears as if Alice would send him canonical messages $|a^n\rangle$ over the basis alphabet and of length $n$, composed from the basis alphabet ${\cal B}_{\cal Q}^n$ with the probability
\begin{eqnarray}
	q(a^n)&=&\langle a^n|\sigma|a^n\rangle		
\end{eqnarray}

\section{Summary and Outlook}

A framework has been worked out that makes the theoretical description of \emph{many-letter states} possible, i.e. states consisting of arbitrary superpositions of quantum messages of distinct length.
The space spanned by these states is the \emph{many-letter space}, which is an infinite direct sum over all \emph{block spaces}, i.e. finite dimensional Hilbert spaces containing quantum messages of fixed length. In the many-letter space a \emph{length operator} is defineable whose eigenspaces are the block spaces and where each eigenvalue is the number of letter systems forming the corresponding eigenspace. 

The concept of many-letter messages can be applied to many topics of quantum information theory. Imagine a source of photons being sent sequentially, but whose number is controlled by the state of a quantum mechanical system. A superposition of input states will result in a superposition of distinguishable photon states of varying number forming a many-letter message whose length is a quantum mechanical observable with distinct values in superposition. Quantum communication, extended to the framework of many-letters, obtains new features. Quantum cryptography might also be affected (imagine an eavesdropper who is not allowed to measure the length of a message without disturbing it), as well as quantum computation (the output of a quantum algorithm can be regarded as a many-letter message). It is also interesting to study the entanglement of many-letter messages. Altogether, I hope that the presented concept will be helpful in many fields of quantum information theory.

\section{Acknowledgements}

I would like to thank Jens Eisert, Timo Felbinger, Alexander Albus, and Shash Virmani for fruitful and intensive discussions about the topic of this paper.

%\bibliography{lossless}

\end{multicols}

\end{document}